\documentstyle[preprint,aps]{revtex}
\begin{document}
\draft
\hyphenation{Nijmegen}
\hyphenation{Rijken}
\hyphenation{San-den}

\title{An accurate nucleon-nucleon potential with charge-independence breaking}

\author{R.B.\ Wiringa}
\address{Physics Division, Argonne National Laboratory,
         Argonne, IL 60439, USA}
\author{V.G.J.\ Stoks}
\address{School of Physical Sciences, The Flinders University of
         South Australia, Bedford Park, South Australia 5042, Australia}
\author{R.\ Schiavilla}
\address{CEBAF Theory Group, Newport News, VA 23606, USA \\
and \\
Department of Physics, Old Dominion University, Norfolk, VA 23529, USA}

\date{August 15, 1994}
\maketitle

\begin{abstract}
We present a new high-quality nucleon-nucleon potential with explicit charge
dependence and charge asymmetry, which we designate Argonne $v_{18}$.
The model has a charge-independent part with fourteen operator
components that is an updated version of the Argonne $v_{14}$ potential.
Three additional charge-dependent and one charge-asymmetric operators
are added, along with a complete electromagnetic interaction.
The potential has been fit directly to the Nijmegen $pp$
and $np$ scattering data base, low-energy $nn$ scattering parameters,
and deuteron binding energy.
With 40 adjustable parameters it gives a $\chi^{2}$ per datum
of 1.09 for 4301 $pp$ and $np$ data in the range 0--350 MeV.
\end{abstract}

\narrowtext

\section{INTRODUCTION}
Traditionally, nucleon-nucleon ($N\!N$) potentials are constructed by
fitting $np$ data for $T=0$ states and either $np$ or $pp$ data for
$T=1$ states. Examples of potentials fit to $np$ data in all states
are the Argonne $v_{14}$~\cite{Wir84}, Urbana $v_{14}$~\cite{Lag81},
and most of the Bonn potentials~\cite{Mac87,Mac89}.
In contrast, the Reid~\cite{Rei68}, Nijmegen~\cite{Nag78}, and
Paris~\cite{Lac80} potentials were fit to $pp$ data for $T=1$ channels.
Unfortunately, potential models which have been fit only to the $np$
data often give a poor description of the $pp$ data~\cite{St93b},
even after applying the necessary corrections for the Coulomb interaction.
By the same token, potentials fit to $pp$ data in $T=1$ states give only
a mediocre description of $np$ data. Fundamentally, this problem is due
to charge-independence breaking in the strong interaction.

In the present work we construct an updated version of the Argonne
potential that fits both $pp$ and $np$ data, as well as low-energy $nn$
scattering parameters and deuteron properties.
The strong interaction potential is written in an operator format that
depends on the values of $S$, $T$, and $T_{z}$ of the $N\!N$ pair.
We then project the potential into a charge-independent (CI) part that has
fourteen operator components (as in the older Argonne $v_{14}$
model) and a charge-independence breaking (CIB) part that
has three charge-dependent (CD) and one charge-asymmetric (CA) operators.
We also include a complete electromagnetic potential, containing Coulomb,
Darwin-Foldy, vacuum polarization, and magnetic moment terms with
finite-size effects. We designate the new model Argonne $v_{18}$.

In a number of applications it is important for an $N\!N$ potential
to reproduce correct $np$ and $pp$ scattering parameters.
For example, in thermal neutron radiative capture on the proton,
$p(n,\gamma)d$, it is crucial to have the correct singlet $np$ scattering
length in the initial state to get the cross section. However, in low-energy
proton weak capture, $p(p,e^{+}\nu_{e})d$, it is equally important that
the correct $pp$ scattering length be provided by the interaction.
Clearly, a complete potential model should meet both requirements.

Another important application is in the formulation of three-nucleon
($N\!N\!N$) potentials. In general, nuclei are underbound using only
$N\!N$ potentials fit to the scattering data. Nontrivial many-nucleon
interactions are expected to make up a portion of the missing binding
energy. Phenomenologically we may choose to construct a many-body
Hamiltonian, such as
\begin{equation}
    H = \sum_{i}\frac{-\hbar^{2}}{2M_{i}}\nabla^{2}_{i}
      + \sum_{i<j}v_{ij} + \sum_{i<j<k}V_{ijk} \ ,  \label{Hamiltonian}
\end{equation}
and constrain the strength parameters of the $N\!N\!N$ potential by
requiring that $H$ gives the correct trinucleon binding energy.
Similar considerations apply if we choose a relativistic formulation.
Clearly, such constraints are ambiguous or even meaningless if the
$N\!N$ potential used in the calculations does not adequately
describe the two-nucleon data.
For ${^{3}}$He (${^{3}}$H), in which the $N\!N$ interaction underbinds by
$\sim$1 MeV, there are two $np$ pairs and one $pp$ ($nn$) pair.
To a good approximation, the two $np$ pairs will be in the $S=1$, $T=0$ state
75\% of the time, and in the $S=0$, $T=1$ state 25\% of the time, while the
$pp$ ($nn$) pair will be pure $S=0$, $T=1$.
If the chosen $N\!N$ potential fits only the more repulsive $pp$ ($nn$)
data in the $T=1$ state, we would get a smaller $N\!N$
contribution to the binding energy and thus overestimate the $N\!N\!N$
potential strength required. By the same token, a model fit to $np$
data in the $T=1$ state would be too attractive and we would
underestimate the $N\!N\!N$ potential. The difference can be as much as
0.4 MeV, leading to variations in the $N\!N\!N$ potential strength of
order $\pm20\%$.  This would have significant effects in larger many-body
systems.

Because we include a complete electromagnetic potential and fit
low-energy $nn$ scattering data, the present model also can be used to
study charge-symmetry breaking, as in the ${^{3}}$H -- ${^{3}}$He
mass difference~\cite{Wu90}, or more generally the Nolen-Schiffer
anomaly~\cite{Nol69}.
The electromagnetic potential is in principle well-known and is the
longest-range part of the interaction. Potential models commonly fit
the deuteron energy to better than 1 keV accuracy. Since we find that
the electromagnetic terms give a non-negligible 18 keV repulsion in
the deuteron and moderate shifts in the $np$ and $nn$ scattering lengths,
we deem it desirable to include these terms explicitly.

The major goal of the present work is to construct a nonrelativistic
potential that can be used easily in nuclear many-body calculations and
that accurately fits both $pp$ and $np$ data.
We adopt the local operator structure of the older Argonne $v_{14}$
and Urbana $v_{14}$ potentials, which have been used extensively
in calculations of finite nuclei, nuclear matter, and neutron
stars~\cite{Wir91,Pie92,Wir88}.
The assumption of an underlying operator structure relates all partial
waves in a simple manner, without imposing a one-boson-exchange (OBE)
form which might be too restrictive at short distances.
Recently, the Nijmegen group has shown~\cite{Sto94} that it is feasible
to construct potential models which fit the $N\!N$ data with the almost
perfect $\chi^{2}$ per datum of 1.
However, these models differ in each partial wave and thus implicitly
introduce nonlocalities from one partial wave to the next that may be
difficult to characterize and treat accurately in many-body calculations.
When they limit the potential to an OBE form, which has a local operator
structure (save for a nonlocal part in the central potential) describing
all partial waves simultaneously, the $\chi^{2}$ per datum increases to
1.87, albeit with a much smaller number of parameters.
The present model is a compromise between these two approaches, adopting
a phenomenological form (unrestricted by an OBE picture) at short distances,
but maintaining a local operator structure.
The potential was directly fit to the Nijmegen $N\!N$ scattering data
base~\cite{Ber90,St93a}, which contains 1787 $pp$ and 2514 $np$ data in
the range 0--350 MeV, and has an excellent $\chi^{2}$ per datum of 1.09.

In Sec.~\ref{sec:form} we present the analytical form of the
potential in the various spin and isospin states. Special attention is
given to the electromagnetic part of the interaction. The free parameters
are fit to the $N\!N$ scattering data and deuteron binding energy
in Sec.~\ref{sec:data}, where we also present the phase shifts.
Section~\ref{sec:projection} discusses the projection of the potential
into operator format. Static deuteron properties and electromagnetic form
factors, with relativisitic and exchange current contributions, are
presented in Sec.~\ref{sec:deuter}.
Conclusions and an outlook are given in Sec.~\ref{sec:outlook}.

\pagebreak

\section{FORM OF THE POTENTIAL \protect\\
         IN \protect\bbox{S, T, T_{\lowercase{z}}} STATES}
\label{sec:form}
The $N\!N$ potential is written as a sum of an electromagnetic (EM) part, a
one-pion-exchange (OPE) part, and an intermediate- and short-range
phenomenological part:
\begin{equation}
   v(N\!N) = v^{EM}(N\!N) + v^{\pi}(N\!N) + v^{R}(N\!N) \ .
\end{equation}
The EM interaction is the same as that used in the Nijmegen
partial-wave analysis, with the addition of short-range terms and
finite-size effects~\cite{Fri76,Aus83,Sto90}.
(Values for the masses and other physical constants used in the following
formulae are given in Table~\ref{constants}.)
For $pp$ scattering we include one- and two-photon Coulomb terms, the
Darwin-Foldy term, vacuum polarization, and the magnetic moment
interaction, each with an appropriate form factor:
\widetext
\begin{equation}
   v^{EM}(pp) = V_{C1}(pp) + V_{C2} + V_{DF} + V_{VP} + V_{MM}(pp) \ .
\end{equation}
Here
\begin{eqnarray}
 &&  V_{C1}(pp) = \alpha'\frac{F_{C}(r)}{r} \ ,     \\
 &&  V_{C2} = -\frac{\alpha}{2M^{2}_{p}}
             \left[(\nabla^{2}+k^{2})\frac{F_{C}(r)}{r}
           + \frac{F_{C}(r)}{r}(\nabla^{2}+k^{2})\right]
          \approx -\frac{\alpha\alpha'}{M_{p}}
                     \left[\frac{F_{C}(r)}{r}\right]^{2} \ ,  \\
 &&  V_{DF} = - \frac{\alpha}{4M^{2}_{p}}F_{\delta}(r) \ ,    \\
 &&  V_{VP} = \frac{2\alpha\alpha'}{3\pi}\frac{F_{C}(r)}{r}
            \int^{\infty}_{1} dx \;
            e^{-2m_{e}rx}\left[1+\frac{1}{2x^{2}}\right]
            \frac{(x^{2}-1)^{1/2}}{x^{2}} \ ,                 \\
 &&  V_{MM}(pp) = -\frac{\alpha}{4M^{2}_{p}}\mu^{2}_{p}
                 \left[\frac{2}{3} F_{\delta}(r)
                 \bbox{\sigma}_{i}\!\cdot\!\bbox{\sigma}_{j}
              + \frac{F_{t}(r)}{r^{3}}S_{ij} \right]
              - \frac{\alpha}{2M^{2}_{p}}(4\mu_{p}-1)
                 \frac{F_{ls}(r)}{r^{3}}{\bf L}\!\cdot\!{\bf S} \ .
\end{eqnarray}
\narrowtext\noindent
The Coulomb interaction includes an energy dependence through the
$\alpha' \equiv 2k\alpha/(M_{p}v_{\rm lab})$~\cite{Bre55}, which is
significantly different from $\alpha$ at even moderate energies
($\sim20\%$ difference at $T_{\rm lab}=250$ MeV).
The vacuum polarization and two-photon Coulomb interaction are important
for fitting the high-precision low-energy scattering data.
The $F_{C}$, $F_{\delta}$, $F_{t}$, and $F_{ls}$ are short-range
functions that represent the finite size of the nucleon charge distributions.
They have been obtained under the assumption that the nucleon
form factors are well represented by a dipole form
\begin{equation}
    G^{p}_{E} = \frac{G^{p}_{M}}{\mu_{p}} = \frac{G^{n}_{M}}{\mu_{n}}
              = G_{D} = \left(1+\frac{q^{2}}{b^{2}}\right)^{-2} \ ,
\end{equation}
where $b=4.27$ fm$^{-1}$. The functions are given by
\begin{eqnarray}
 &&   F_{C}(r) = 1 - \left(\! 1 + \frac{11}{16}x + \frac{3}{16}x^{2}
                   + \frac{1}{48}x^{3}\right)e^{-x}, \nonumber\\
 &&   F_{\delta}(r) = b^{3}\left(\frac{1}{16} + \frac{1}{16}x
                      + \frac{1}{48}x^{2}\right)e^{-x}, \nonumber\\
 &&   F_{t}(r) = 1 - \left(\! 1 + x + \frac{1}{2}x^{2} + \frac{1}{6}x^{3}
                   + \frac{1}{24}x^{4} + \frac{1}{144}x^{5}
                     \right)e^{-x}, \nonumber\\
 &&   F_{ls}(r) = 1 - \left(\! 1 + x + \frac{1}{2}x^{2} + \frac{7}{48}x^{3}
                    + \frac{1}{48}x^{4}\right)e^{-x}.
\end{eqnarray}
with $x=br$. The derivation of $F_{C}$ is given in ref.~\cite{Aue72}, while
the others are related by $F_{\delta}=-\nabla^{2}(F_{C}/r)$,
$F_{t}=(F_{C}/r)''-(F_{C}/r)'/r$, and $F_{ls}=(F_{C}/r)'/r$.
In the limit of point nucleons, $F_{C}=F_{t}=F_{ls}=1$ and
$F_{\delta}=4\pi\delta^{3}({\bf r})$.
These form factors are illustrated in Fig.~\ref{figFFs}.
The use of $F_{C}$ in $V_{VP}$ is an approximate method of removing the
$1/r$ singularity (the logarithmic singularity remains) which is justified
by its short-range and the overall smallness of the term.
Similarly, the use of $F_{C}^{2}$ in $V_{C2}$ is an approximate method
of removing the $1/r^{2}$ singularity.
We note that because we use the Sachs nucleon form factors,
there are no additional magnetic Darwin-Foldy terms~\cite{Fri75}.

For the $np$ system we include a Coulomb term attributable to the neutron
charge distribution in addition to the interaction between magnetic moments,
\begin{equation}
   v^{EM}(np) = V_{C1}(np) + V_{MM}(np) \ .
\end{equation}
Here
\begin{equation}
   V_{C1}(np) = \alpha \beta_{n} \frac{F_{np}(r)}{r} \ ,
\end{equation}
where the function $F_{np}$ is obtained assuming the neutron electric form
factor~\cite{Fri75}
\begin{equation}
    G^{n}_{E} = \beta_{n} q^{2} \left(1+\frac{q^{2}}{b^{2}}\right)^{-3} \ .
\end{equation}
Here $\beta_{n}\equiv [dG^{n}_{E}/dq^{2}]_{q=0} = 0.0189$ fm$^{2}$,
the experimentally measured slope~\cite{Kro73}.
We have checked this form factor in a self-consistent calculation of the
deuteron structure function $A(q^{2})$ used to extract $G^{n}_{E}$~\cite{Pla90}
and find it gives a fairly good fit to the data.
This simple form leads to
\begin{equation}
    F_{np}(r) = b^{2}\left( 15x + 15x^{2} + 6x^{3} + x^{4}\right)
                \frac{e^{-x}}{384} \ .
\end{equation}
The $F_{np}$ is also shown in Fig.~\ref{figFFs}.
The magnetic moment interaction is given by
\begin{eqnarray}
   V_{MM}(np) &=& -\frac{\alpha}{4M_{n}M_{p}}\mu_{n}\mu_{p}
                 \left[\frac{2}{3}
                 F_{\delta}(r)\bbox{\sigma}_{i}\!\cdot\!\bbox{\sigma}_{j}
              + \frac{F_{t}(r)}{r^{3}}S_{ij} \right]         \nonumber\\
              && - \frac{\alpha}{2M_{n}M_{r}}\mu_{n}
                \frac{F_{ls}(r)}{r^{3}}
                  ({\bf L}\!\cdot\!{\bf S}+{\bf L}\!\cdot\!{\bf A}) \ ,
\end{eqnarray}
where $M_{r}$ is the nucleon reduced mass. The term proportional to
${\bf A}=\frac{1}{2}(\bbox{\sigma}_{i}-\bbox{\sigma}_{j})$
is a ``class IV'' charge-asymmetric force~\cite{Hen79}, which mixes
spin-singlet and spin-triplet states. Its contribution is very small,
and we only include it when we construct the magnetic moment scattering
amplitude~\cite{Sto90}.

Finally, for $nn$ scattering, we neglect the Coulomb interaction between
the neutron form factors, so there is only a magnetic moment term
\begin{eqnarray}
   v^{EM}(nn) &=& V_{MM}(nn)                               \nonumber\\
              &=& -\frac{\alpha}{4M^{2}_{n}}\mu^{2}_{n}
                 \left[\frac{2}{3}
                 F_{\delta}(r)\bbox{\sigma}_{i}\!\cdot\!\bbox{\sigma}_{j}
              + \frac{F_{t}(r)}{r^{3}}S_{ij} \right] \ .
\end{eqnarray}

The charge-dependent structure of the OPE potential is the same as that
used in the Nijmegen partial-wave analysis and reads
\begin{eqnarray}
   v^{\pi}(pp) &=& f_{pp}^{2} v_{\pi}(m_{\pi^{0}}) \ , \nonumber\\
   v^{\pi}(np) &=& f_{pp}f_{nn} v_{\pi}(m_{\pi^{0}})
      + (-)^{T+1}2f_{c}^{2} v_{\pi}(m_{\pi^{\pm}})      \ , \label{Vope}\\
   v^{\pi}(nn) &=& f_{nn}^{2} v_{\pi}(m_{\pi^{0}}) \ , \nonumber
\end{eqnarray}
where $T$ is the isospin and
\begin{equation}
   v_{\pi}(m)=\left(\frac{m}{m_{s}}\right)^{2}{\textstyle\frac{1}{3}}
              mc^{2} \left[
              Y_{\mu}(r)\bbox{\sigma}_{i}\!\cdot\!\bbox{\sigma}_{j} +
              T_{\mu}(r) S_{ij}  \right] \ .      \label{Vpi}
\end{equation}
(Strictly speaking, the neutron-proton mass difference gives rise to an
OPE ``class IV'' force as well, which again we only explicitly include
when we construct the OPE scattering amplitude~\cite{Sto90}.)
Here $Y_{\mu}(r)$ and $T_{\mu}(r)$ are the usual Yukawa and tensor functions
with the exponential cutoff of the Urbana and Argonne $v_{14}$ models
\begin{eqnarray}
   Y_{\mu}(r)&=&\frac{e^{-\mu r}}{\mu r}
              \left(1-e^{-cr^{2}}\right)          \ , \nonumber\\
   T_{\mu}(r)&=&\left(1+\frac{3}{\mu r}+\frac{3}{(\mu r)^{2}}\right)
              \frac{e^{-\mu r}}{\mu r} \left(1-e^{-cr^{2}}\right)^{2} \ ,
              \label{Tmu}
\end{eqnarray}
where $\mu=mc/\hbar$.
The scaling mass $m_{s}$, introduced in Eq.~(\ref{Vpi}) to make the
coupling constant dimensionless, is taken to be the charged-pion mass,
$m_{\pi^{\pm}}$. The Nijmegen partial-wave analysis of $N\!N$ scattering
data below 350 MeV finds very little difference between the coupling
constants~\cite{St93c}, so we choose them to be charge-independent, i.e.,
$f_{pp}=-f_{nn}=f_{c}\equiv f$, with the recommended value $f^{2}=0.075$.
Thus all charge dependence in Eqs.~(\ref{Vope}) is due simply to the
difference in the charged- and neutral-pion masses.

The remaining intermediate- and short-range phenomenological part of the
potential is expressed, as in the Argonne $v_{14}$ model, as a sum of central,
$L^{2}$, tensor, spin-orbit, and quadratic spin-orbit terms (abbreviated as
$c, l2, t, ls, ls2$, respectively) in different $S$, $T$, and $T_{z}$ states:
\widetext
\begin{equation}
   v^{R}_{ST}(N\!N) = v^{c}_{ST,N\!N}(r) + v^{l2}_{ST,N\!N}(r)L^{2}
      + v^{t}_{ST,N\!N}(r)S_{12}
      + v^{ls}_{ST,N\!N}(r){\bf L}\!\cdot\!{\bf S}
      + v^{ls2}_{ST,N\!N}(r)({\bf L}\!\cdot\!{\bf S})^{2} \ .
\end{equation}
Each of these terms is given the general form
\begin{equation}
    v^{i}_{ST,N\!N}(r) = I^{i}_{ST,N\!N} T_{\mu}^{2}(r)
       + \left[P^{i}_{ST,N\!N} + \mu r\,Q^{i}_{ST,N\!N}
       + (\mu r)^{2}R^{i}_{ST,N\!N}\right] W(r) \ ,
\end{equation}
\narrowtext\noindent
where $\mu=\textstyle\frac{1}{3}(m_{\pi^{0}}+2m_{\pi^{\pm}})c/\hbar$ is
the average of the pion masses and $T_{\mu}(r)$ is given by Eq.~(\ref{Tmu}).
Thus the $T_{\mu}^{2}(r)$ term has the range of a two-pion-exchange force.
The $W(r)$ is a Woods-Saxon function which provides the short-range core:
\begin{equation}
      W(r)=\left[1+e^{(r-r_{0})/a}\right]^{-1} \ .   \label{WS}
\end{equation}
The four sets of constants $I^{i}_{ST,N\!N}$, $P^{i}_{ST,N\!N}$,
$Q^{i}_{ST,N\!N}$, and $R^{i}_{ST,N\!N}$ are parameters to be fit to data.
However, we also impose a regularization condition at the origin which
reduces the number of free parameters by one for each $v^{i}_{ST,N\!N}$.
We require that
\begin{eqnarray}
    v^{t}_{ST,N\!N}(r=0) &=& 0          \ , \nonumber\\
    \frac{\partial v^{i\neq t}_{ST,N\!N}}{\partial r} |_{r=0} &=& 0 \ .
                                            \label{constraint}
\end{eqnarray}
Since the tensor part of the OPE potential already vanishes at $r=0$,
the first condition is satisfied by setting $P^{t}_{ST,N\!N}=0$.
The second condition is equivalent to fixing, for $i\neq t$,
\begin{equation}
   Q^{i}_{ST,N\!N}=-\frac{1}{\mu W(0)}\left[P^{i}_{ST,N\!N}
     \frac{\partial W}{\partial r} + \delta_{ic}
     \frac{\partial v^{\pi}_{ST}}{\partial r} \right]_{r=0}  \ ,
\end{equation}
where we only have to evaluate the derivative of the spin-spin part
of the OPE potential.

\section{DATA FITTING}
\label{sec:data}
An initial survey of possible potential forms was made by fitting to
the $\chi^{2}$ hypersurface of the Nijmegen partial-wave analysis of
$pp$ and $np$ data~\cite{St93a}. These studies helped select the final
form of the potential ($\sim$10 variations were tried) and the values
of the function shape parameters $c$, $r_{0}$, and $a$. Eventually,
the cutoff parameter in the OPE functions $Y_{\mu}(r)$ and $T_{\mu}(r)$
was set at $c=2.1$ fm$^{-2}$, while the parameters in the short-range
Woods-Saxon $W(r)$ were set at $r_{0}=0.5$ fm and $a=0.2$ fm. This value
of $c$ is slightly different from the 2.0 fm$^{-2}$ used in the Urbana
and Argonne $v_{14}$ models, while $r_{0}$ and $a$ are the same.
Attempts to make a softer-core model led to a poorer fit.
Sensitivity to the OPE coupling constant was also checked before the
recommended value~\cite{St93c}, $f^{2}=0.075$, was adopted as optimal.

Once these four parameters were set, a preliminary fit of the remaining
parameters $I^{i}_{ST,N\!N}, P^{i\neq t}_{ST,N\!N}, Q^{t}_{ST,N\!N}$,
and $R^{i}_{ST,N\!N}$ to the phase shifts was made. The final values
were obtained by a direct fit to the Nijmegen $pp$ and $np$ scattering
data base and the deuteron binding energy.
We use nonrelativistic kinematics, i.e., the deuteron binding energy is
taken as $E_{d}=\kappa^{2}/2M_{\rm r}$. In practice, we found no benefit
to including an $R^{i}_{ST,N\!N}$ in spin-singlet states, so these
values were set to zero. Also, we found no indication of a need for
charge dependence in the phenomenological part of spin-triplet states.
In the final fit there are 40 nonzero intermediate- and short-range
parameters.
In addition, we fit the $nn$ scattering length and effective range
as determined by $d(\pi^{-},\gamma)nn$ experiments~\cite{Gab79}, which
seem most consistent with charge-symmetry breaking in the trinucleon. This
is done by slightly modifying the short-range $S=0$, $T=1$ $pp$ potential,
resulting in a small difference between $P^{c}_{01,pp}$ and $P^{c}_{01,nn}$.
The same difference was carried over to the $P^{c}_{11,nn}$ parameter,
as discussed below in Sec.~\ref{sec:projection}.
All these parameter values are given in Table~\ref{parameters}.

The Nijmegen $N\!N$ scattering data base~\cite{Ber90,St93a} includes
1787 $pp$ data (1656 observables and 131 normalization data) and 2514
$np$ data (2366 observables and 148 normalization data) in the range
0--350 MeV.
The total $\chi^{2}$ of the potential is 4675, divided into
$\chi^{2}(pp)=1962$ and $\chi^{2}(np)=2713$.
A detailed breakdown of the $\chi^{2}$, analogous to Ref.~\cite{St93b}
for the $pp$ data, is given in Table~\ref{chi2}. We also show the
breakdown for the Nijmegen combined partial-wave analysis~\cite{St93a}.
The difference between these starts to increase beyond $\sim$150 MeV.
We should mention that there are a number of groups of $np$ total cross
section data which extend over a wide energy range. So in order to
present the results in the form of Table~\ref{chi2}, we had to split
each of these groups into a number of subgroups, each contributing in
its appropriate energy bin. Whenever one of these groups has a
normalization error, we choose to apply this same normalization for
each of its subgroups. As a consequence, the number of $np$ data in
Table~\ref{chi2} is increased by 12, while the total $\chi^{2}(np)$ is
lowered by 28. The reason for this reduction in $\chi^{2}$ is that these
12 extra normalizations are optimized for each subgroup separately.

As an independent test, we have also checked our results with the
scattering analysis interactive dial-in (SAID) program, of the Virginia
Polytechnic Institute and State University (VPI\&SU) group~\cite{Arn92}.
We provided SAID with phase shifts calculated at the seventeen energies
1, 5, 10, 25 (25) 350 MeV. The SAID program then uses an interpolation
routine to provide the phase shifts at all energies at which the
experimental data were measured. For the $S$ waves below 25 MeV, this
interpolation deviates slightly from the actual values of the potential.
Moreover, the treatment of the electromagnetic contributions is less
sophisticated in the VPI\&SU analysis, which leads to a large discrepancy
for laboratory energies below 2 MeV. When we then compare with the data
between 2 and 350 MeV, we obtain a $\chi^{2}(pp)=2107$ for 1644 $pp$
data and a $\chi^{2}(np)=4157$ for 3020 $np$ data, all from their data
set NN943. The total $\chi^{2}$ per datum by this comparison is still
a very respectable 1.34.

To demonstrate once more the importance of fitting to both $pp$ and $np$
data, we used the $np$ version of the Argonne $v_{18}$ potential,
included the proper electromagnetic interaction, and confronted
it with the $pp$ data. In this way we partially simulate the comparison
with the $pp$ data of a potential model fit only to the $np$ data.
Of course, the analogue is not perfect, because in the Argonne $v_{18}$
potential the $l2,t,ls,ls2$ parameters in the triplet $T=1$ partial
waves are fit to both $pp$ and $np$ data. Still, this modified $np$
potential gives a $\chi^{2}$ per datum of 4.4 for the $pp$ data between
2 and 350 MeV, which is much worse than the $\chi^{2}$ per datum of 1.1
for the actual $pp$ Argonne $v_{18}$ potential on the same energy interval.
Similarly, we can replace the $T=1$ $np$ part of the Argonne $v_{18}$
potential by the $T=1$ $pp$ part and confront it with the $np$ data.
This modified potential then gives a $\chi^{2}$ per datum of 1.8 on the
$np$ data between 2 and 350 MeV, rather than 1.1 for the actual $np$
Argonne $v_{18}$.

The $L=0$ phase shifts are calculated using the potentials discussed
in Sec.~\ref{sec:form}, i.e., including the complete electromagnetic
interaction, and by matching to electromagnetic wave functions
($\delta^{EM}_{EM+N}$ in the notation of Ref.~\cite{Ber88}).
For $L\neq0$ we use the fact that $\delta^{EM}_{EM+N}$ can be
reasonably approximated~\cite{Ber88,St93a} by only including the
Coulomb interaction with $\alpha'$ (in case of $pp$ scattering) or no
electromagnetic interaction at all (in the case of $np$ or $nn$
scattering).
The resulting phase shifts for partial waves with $J\leq3$ are shown in
Table~\ref{phspp} for $pp$ scattering, in Table~\ref{phsnn} for $nn$,
in Table~\ref{phsnp1} for $np$ in $T=1$ states, and in Table~\ref{phsnp0}
for $np$ in $T=0$ states.
Note that the non-$S$ $pp$ phase shifts in Table~\ref{phspp} are calculated
including the form factor $F_{C}(r)$ in the Coulomb potential $V_{C1}(pp)$.

In addition, we show figures of some of the more interesting phases and
compare to the Nijmegen multienergy partial-wave analysis~\cite{St93a},
the single-energy analysis from SAID~\cite{Arn92}, and recent single-energy
analyses by Bugg and Bryan~\cite{Bug92}, and by Henneck~\cite{Hen93}.
In Fig.~\ref{fig1S0} the $pp$, $nn$, and $np$ $^{1}S_{0}$ phases of
Argonne $v_{18}$ are shown, and seen to be in good agreement with
the various analyses.
The charge dependence is clearly evident; a discussion of the relative
size of various contributions to charge-independence breaking is given
below in Sec.~\ref{sec:projection}.
The $^{3}P_{0}$ phases are shown in Fig.~\ref{fig3P0}; the $^{3}P_{0}$
channel displays the second greatest amount of CIB after the $^{1}S_{0}$
channel.
Again there seems to be reasonable agreement with the various
partial-wave analyses.
The $\epsilon_{1}$ mixing parameter, shown in Fig.~\ref{figeps1}, is
both the most difficult to determine in single-energy analyses, as
indicated by the range of values and size of error bars, and one of the
most important because of its relation to the strength of the tensor
interaction. The Argonne $v_{18}$ value tracks the Nijmegen multienergy
analysis up to $T_{\rm lab}=100$ MeV before deviating slightly on the
high side. However, the differences with the Nijmegen multienergy analysis
are still within two standard deviations.
Finally, the $^{1}P_{1}$ phase shift, which is intimately related to
the $\epsilon_{1}$ mixing parameter, is shown in Fig.~\ref{fig1P1}.
Here the present model is somewhat less repulsive than the various
partial-wave analyses above 150 MeV.

The low-energy scattering parameters are shown in Table~\ref{lowpar}
and compared to experimental results~\cite{Ber88,Gab79,Koe75}.
The scattering lengths and effective ranges are calculated both with
and without the electromagnetic interaction. Without the electromagnetic
interaction, the effective range function is simply given by
$F(k^{2})=k\cot\delta_{N}=-1/a+{\textstyle\frac{1}{2}}rk^{2}+{\cal O}(k^{4})$.
In the presence of the electromagnetic interaction, we have to use
a more complicated effective range function~\cite{Ber88}, where the
phase shifts are with respect to the full long-range electromagnetic
interaction.

\section{PROJECTION INTO OPERATOR FORMAT}
\label{sec:projection}
We can project the strong interaction potential given above from
$S, T, T_{z}$ states into an operator format with 18 terms
\begin{equation}
       v_{ij}=\sum_{p=1,18} v_{p}(r_{ij}) O^{p}_{ij} \ .
\end{equation}
Here the first fourteen operators are the same charge-independent ones
used in the Argonne $v_{14}$ potential and are given by
\widetext
\begin{eqnarray}
   O^{p=1,14}_{ij} &=& 1, \bbox{\tau}_{i}\!\cdot\!\bbox{\tau}_{j},\,
       \bbox{\sigma}_{i}\!\cdot\!\bbox{\sigma}_{j},
       (\bbox{\sigma}_{i}\!\cdot\!\bbox{\sigma}_{j})
       (\bbox{\tau}_{i}\!\cdot\!\bbox{\tau}_{j}),\,
    S_{ij}, S_{ij}(\bbox{\tau}_{i}\!\cdot\!\bbox{\tau}_{j}),\,
    {\bf L}\!\cdot\!{\bf S}, {\bf L}\!\cdot\!{\bf S}
        (\bbox{\tau}_{i}\!\cdot\!\bbox{\tau}_{j}),         \nonumber\\
  & & L^{2}, L^{2}(\bbox{\tau}_{i}\!\cdot\!\bbox{\tau}_{j}),\,
    L^{2}(\bbox{\sigma}_{i}\!\cdot\!\bbox{\sigma}_{j}),
    L^{2}(\bbox{\sigma}_{i}\!\cdot\!\bbox{\sigma}_{j})
         (\bbox{\tau}_{i}\!\cdot\!\bbox{\tau}_{j}),\,
    ({\bf L}\!\cdot\!{\bf S})^{2}, ({\bf L}\!\cdot\!{\bf S})^{2}
        (\bbox{\tau}_{i}\!\cdot\!\bbox{\tau}_{j})\ .
\end{eqnarray}
\narrowtext\noindent
These fourteen components are denoted by the abbreviations $c$, $\tau$,
$\sigma$, $\sigma\tau$, $t$, $t\tau$, $ls$, $ls\tau$, $l2$, $l2\tau$,
$l2\sigma$, $l2\sigma\tau$, $ls2$, and $ls2\tau$.
The four additional operators break charge independence and are given by
\begin{equation}
   O^{p=15,18}_{ij}  =  T_{ij}, \,
        (\bbox{\sigma}_{i}\!\cdot\!\bbox{\sigma}_{j})
        T_{ij},\, S_{ij}T_{ij},\, (\tau_{zi}+\tau_{zj})\ ,
\end{equation}
where $T_{ij}=3\tau_{zi}\tau_{zj}-\bbox{\tau}_{i}\!\cdot\!\bbox{\tau}_{j}$,
is the isotensor operator, defined analogous to the $S_{ij}$ operator.
These terms are abbreviated as $T$, $\sigma T$, $tT$, and $\tau z$.
The $T$, $\sigma T$, and $tT$ operators are charge-dependent and are
``class II'' forces, while the $\tau z$ operator is charge-asymmetric
and is a ``class III'' force~\cite{Hen79}.

The operator potential terms, $v_{p}$, can be obtained from the channel
potentials, $v^{x}_{ST,N\!N}$, by a simple set of projections.
We first introduce charge splitting for the central $T=1$ states,
\begin{equation}
   v^{c}_{S1,N\!N}=v^{\rm ci}_{S1}+v^{\rm cd}_{S1}T_{ij}
                  +v^{\rm ca}_{S1}(\tau_{zi}+\tau_{zj})  \ . \label{split}
\end{equation}
For the charge-independent potential this implies
\begin{equation}
   v^{\rm ci}_{S1}={\textstyle\frac{1}{3}}(v^{c}_{S1,pp}+v^{c}_{S1,nn}
                                     +v^{c}_{S1,np}) \ .
\end{equation}
We then project
\begin{mathletters}
\begin{eqnarray}
   v_{c}     &=&{\textstyle\frac{1}{16}}(9v^{\rm ci}_{11}
              +3v^{\rm ci}_{10}+3v^{\rm ci}_{01}+ v^{\rm ci}_{00}) \ , \\
   v_{\tau}  &=&{\textstyle\frac{1}{16}}(3v^{\rm ci}_{11}
              -3v^{\rm ci}_{10}+ v^{\rm ci}_{01}- v^{\rm ci}_{00}) \ , \\
   v_{\sigma}&=&{\textstyle\frac{1}{16}}(3v^{\rm ci}_{11}
              + v^{\rm ci}_{10}-3v^{\rm ci}_{01}- v^{\rm ci}_{00}) \ , \\
   v_{\sigma\tau}&=&{\textstyle\frac{1}{16}}(v^{\rm ci}_{11}
              - v^{\rm ci}_{10}- v^{\rm ci}_{01}+ v^{\rm ci}_{00}) \ ,
\end{eqnarray}
\end{mathletters}
where of course $v^{\rm ci}_{10}=v^{c}_{10,np}$ and
$v^{\rm ci}_{00}=v^{c}_{00,np}$.
A similar set of projections is used for the $L^{2}$ parts of the
interaction. For the tensor, spin-orbit, and quadratic spin-orbit pieces,
which exist only in $S=1$ channels, the projections are ($x=t, ls, ls2$)
\begin{mathletters}
\begin{eqnarray}
    v_{x}    &=&{\textstyle\frac{1}{4}}(3v^{x}_{11}+v^{x}_{10}) \ , \\
    v_{x\tau}&=&{\textstyle\frac{1}{4}}( v^{x}_{11}-v^{x}_{10}) \ .
\end{eqnarray}
\end{mathletters}
The charge-dependent terms in Eq.~(\ref{split}) are given by
\begin{equation}
    v^{\rm cd}_{S1}={\textstyle\frac{1}{6}}[{\textstyle\frac{1}{2}}
                      (v^{c}_{S1,pp}+v^{c}_{S1,nn})-v^{c}_{S1,np}] \ ,
\end{equation}
which can be projected as
\begin{mathletters}
\begin{eqnarray}
    v_{T}       &=&{\textstyle\frac{1}{4}}
                   (3v^{\rm cd}_{11}+v^{\rm cd}_{01}) \ , \\
    v_{\sigma T}&=&{\textstyle\frac{1}{4}}
                   ( v^{\rm cd}_{11}-v^{\rm cd}_{01}) \ .
\end{eqnarray}
\end{mathletters}
The charge-dependent tensor term comes only from the spin-triplet
channel, and reads
\begin{equation}
    v_{tT}={\textstyle\frac{1}{6}}[{\textstyle\frac{1}{2}}
           (v^{t}_{11,pp}+v^{t}_{11,nn})-v^{t}_{11,np}] \ .
\end{equation}
Finally, the charge-asymmetric terms are given by
\begin{equation}
    v^{\rm ca}_{S1}={\textstyle\frac{1}{4}}
                      (v^{c}_{S1,pp}-v^{c}_{S1,nn}) \ ,
\end{equation}
which leads to
\begin{mathletters}
\begin{eqnarray}
    v_{\tau z}      &=&{\textstyle\frac{1}{4}}
                       (3v^{\rm ca}_{11}+v^{\rm ca}_{01}) \ , \\
    v_{\sigma\tau z}&=&{\textstyle\frac{1}{4}}
                       ( v^{\rm ca}_{11}-v^{\rm ca}_{01}) \ .
\end{eqnarray}
\end{mathletters}
As discussed in the previous section, we fix $v^{\rm ca}_{01}$ to reproduce
the singlet $nn$ scattering length by adjusting the parameter $P^{c}_{01,nn}$
to be slightly different from $P^{c}_{01,pp}$.
We are unaware of any $nn$ data that would allow us to fix
$v^{\rm ca}_{11}$, but there have been numerous theoretical predictions
for charge-symmetry breaking based on $\rho$-$\omega$ and
$\pi$-$\eta$-$\eta'$ mixing. Such models suggest that $v^{\rm ca}_{11}$
should be somewhat larger than $v^{\rm ca}_{01}$, but with a similar
shape~\cite{Lan82}. In the present work we make the simple assumption
$v^{\rm ca}_{11}=v^{\rm ca}_{01}$, which implies there is no
$v_{\sigma\tau z}$ term. We also neglect the possibility of a
charge-asymmetric tensor term $v_{t\tau z}$, which is why we end up
with only one charge-asymmetric operator in our model.
These choices are reflected in the parameters of Table~\ref{parameters}.

The first four operator components of the potential are shown in
Fig.~\ref{figvc}. The tensor components
are shown in Fig.~\ref{figvt} where we also show the CI part of the
OPE potential used here, and for comparison an OPE potential constructed
using the same coupling constant and a dipole form factor (monopole at
each nucleon-nucleon-pion vertex) with the cutoff mass $\Lambda=900$ MeV.
The spin-orbit and quadratic spin-orbit terms are shown in
Fig.~\ref{figvls}, while the various $L^{2}$ components are shown in
Fig.~\ref{figvl2}. Finally, the charge-dependent and charge-asymmetric
terms are shown in Fig.~\ref{figvcd}, along with the static Coulomb
potential for comparison.

The relative importance of the different CIB components is illustrated
in Table~\ref{phsevol}, where the evolution from the CI part of the
interaction to the full $pp$ interaction is displayed. The successive
columns give the $^{1}S_{0}$ phase shifts for 1) the CI potential with an
average nucleon and average pion mass, 2) with the correct proton mass,
3) with the correct CD OPE tail (i.e., correct neutral-pion mass) but
the CI core, 4) with both the CD OPE and core interactions,
and 5) with the electromagnetic potential added.
{}From these it can be seen that the nucleon mass has a relatively small
effect,
while the CD OPE and core terms have relatively large effects at low
energy, and the core contribution becomes dominant at higher energies.

\section{DEUTERON PROPERTIES}
\label{sec:deuter}
The static deuteron properties are shown in Table~\ref{deuteron} and
compared to experimental values~\cite{Leu82,Eri83,Rod90,Sim81,Lin65,Bis79}.
The binding energy, $E_{d}$, is fit exactly by construction.  The expectation
values for the kinetic energy, $T$, and for the EM, OPE, and remaining
potentials are also shown.  We note that the OPE potential dominates, while
the EM potential gives a small but non-negligible 18 keV contribution,
mostly from the magnetic moment term.
The asymptotic $S$-state normalization, $A_{S}$, the $D/S$-ratio, $\eta$,
and the deuteron radius, $r_{d}$, all come out close to the experimental
values.
The magnetic moment, $\mu_{d}$, and the quadrupole moment, $Q_{d}$, are
both underpredicted in impulse approximation; both have significant
relativistic and meson-exchange corrections, as discussed below.
Finally, the $D$-state percentage is about 5\% smaller than that of the
older Argonne $v_{14}$ model~\cite{Wir84} and almost identical to that of
the Paris potential~\cite{Lac80}.

The $S$- and $D$-wave components of the deuteron wave function are shown
in Fig.~\ref{figuw}, where they are compare to those for the older $v_{14}$
model.  The short-range behavior of the wave function components is
moderately different.  The $A(q^2)$ and $B(q^2)$ structure functions and
tensor polarization $T_{20}(q^2)$ obtained with the present interaction
model are displayed in Figs.~\ref{figA}--\ref{figT20}; the experimental
data is from
Refs.~\cite{Pla90,Sim81,Cra85,Arn75,Auf85,Arn87,Sch84,The91,Dmi85,Gil90}.
The model for the isoscalar electromagnetic current operator has been
discussed in detail in Refs.~\cite{Sch89,Sch91}, here we only summarize its
general structure, which consists of one- and two-body parts. The one-body
part has the standard impulse approximation (IA) form, with inclusion,
in the charge component, of the Darwin-Foldy and spin-orbit relativistic
corrections~\cite{Sch90}. The two-body charge operators contain
contributions that correspond (in an OBE picture) to those obtained from
pion- and vector-meson ($\rho$ and $\omega$) exchanges.
These are obtained from the nonrelativistic reduction of the
Born terms in the corresponding relativistic photoproduction
amplitudes~\cite{Sch90}. The two-body current operators are constructed
from the spin-orbit and quadratic momentum-dependent components of the
interaction with the methods developed in Refs.~\cite{Sch89,Car90}.
We also consider the two-body charge and current operators associated
with the $\rho\pi\gamma$ mechanism. In particular, we include in the
nonrelativistic reduction of its current component the next to leading
order correction arising from the tensor coupling of the $\rho$ meson
to the nucleon~\cite{Sch91}. The H\"ohler parametrization 8.2~\cite{Hoh76} is
used for the electromagnetic form factors of the nucleon, while an
$\omega$-pole term form factor is included at the $\rho\pi\gamma$
electromagnetic vertex.

The calculated $A(q^2)$ structure function is in excellent agreement with
the experimental data over the whole range of measured momentum
transfers. The Darwin-Foldy and spin-orbit relativistic corrections
to the single-nucleon charge operator as well as the leading two-body
charge contribution due to pion exchange play an important role, as it
is evident from Fig.~\ref{figA}. However, these same contributions lead
to a significant discrepancy between theory and experiment in the tensor
polarization. This observable and the $A(q^2)$ structure function are mostly
sensitive to the charge and quadrupole form factors. In particular, the
momentum transfer at which the minimum of $T_{20}(q^2)$ occurs is related
to the position of the charge form factor zero. The relative shift
between the predicted and experimental $T_{20}(q^2)$ minima implies,
therefore, a corresponding shift between the charge form factor zeros.

The calculated $B(q^2)$ structure function is found to overpredict the
experimental data in the momentum transfer range 10--45 fm$^{-2}$, and
has a zero around 60 fm$^{-2}$. The leading two-body contributions
are those due to the spin-orbit and quadratic spin-orbit components of
the interaction. They are of opposite sign. However, the overestimate
of the data indicates that the degree of cancellation between them is
not quite enough. The $\rho\pi\gamma$ current contribution is small
over the momentum transfer range considered here (we have used the rather
soft cutoff values of 0.75 GeV and 1.25 GeV at the $\pi N\!N$ and
$\rho N\!N$ vertices, respectively, as suggested in Ref.~\cite{Car91}.)
At present, the two-body currents associated with the quadratic
spin-orbit and $L^2$ components of the interaction are essentially
obtained by minimal substitution ${\bf p}_i \rightarrow {\bf p}_i -
[G_E^s(q^2)+G_E^v(q^2) \tau_{z,i}] {\bf A}({\bf r}_i)$, where ${\bf A}$ is
the vector potential, $G_E^s$ and $G_E^v$ the isoscalar and isovector
nucleon electric form factors~\cite{Sch89}. It would be desirable to
construct these current components in a more systematic way, as suggested
in Ref.~\cite{Blu92}.

Finally, the values for the quadrupole and magnetic moments obtained
with the full charge and current operators are: $Q_d=0.275$ fm$^2$ and
$\mu_{d}=0.871$ $\mu_{o}$. The measured quadrupole moment
is underestimated by roughly 4\%, while the measured magnetic
moment is overestimated by 1.5\%. The two-body charge and
current contributions amount to 2\% and 3\% increases of
the IA values for $Q_{d}$ and $\mu_{d}$, as listed in Table~\ref{deuteron}.

\section{CONCLUSIONS AND OUTLOOK}
\label{sec:outlook}
We have constructed a nonrelativistic $N\!N$ potential with a local
operator structure that gives an excellent fit to $pp$ and $np$
scattering data, as well as to low-energy $nn$ scattering and the
deuteron binding energy. We have projected the potential into
charge-independent, charge-dependent, and charge-asymmetric pieces.
In $T=0$ many-body systems only the CI part of the potential will
contribute, while the CA part will contribute in systems with
$T\geq\frac{1}{2}$ and the CD part for $T\geq 1$ systems.
Because of the isotensor projection, the CI part automatically has the
correct average of $\frac{2}{3}$ $pp$ (or $nn$) and $\frac{1}{3}$ $np$
$T=1$ interaction in the trinucleons, thus serving as a correct
reference point for building $N\!N\!N$ potentials.
The CA part will contribute to the energy differences of mirror nuclei
(the Nolen-Schiffer anomaly~\cite{Nol69}), while the CD part will
contribute to the splitting of isobaric analog states, e.g., in the
$A=6$ nuclei. Studies of these effects are in progress.

We have also computed the deuteron electromagnetic properties in both
impulse approximation and with relativistic and exchange-current
corrections. The least satisfactory prediction of the potential is the small
value for $Q_d$, even after corrections are added. The full 4\% discrepancy
between the predicted and empirical $Q_d$ values is unlikely to be
resolved by additional relativistic and/or two-body corrections not
included in the present calculation. Similar low values were found by
the Nijmegen group in their recent fits~\cite{Sto94}, which used rather
different potential forms than the present model. We are less concerned
about the 1.5\% error in the magnetic moment or the overprediction of
the $B(q^{2})$ structure function because of the uncertainties in the
exchange currents discussed above.
The $A(q^{2})$ structure function is very well reproduced, while the
experimental tensor polarization $T_{20}(q^{2})$ still has rather large
error bars above 10 fm${^{-2}}$.

Compared to the older Argonne $v_{14}$ potential, the present model has
a weaker tensor force, which will generally lead to more binding in light
nuclei and less rapid saturation in nuclear matter. This is counteracted
by the weaker attraction in $T=1$ $N\!N$ states because of the mix of
$pp$ and $np$ components. Initial calculations of few-body nuclei with
the $v_{18}$ model show a slight net reduction in the binding energies
of ${^{3}}$H and ${^{4}}$He compared to the $v_{14}$ model.
Another feature of the new model is a moderately greater attraction in
$P$ waves. Few-body nuclei are not sensitive to this part of the $N\!N$
interaction. However, preliminary calculations~\cite{Pie94} of the
binding energy of ${^{16}}$O show a significant improvement in the
relative stability of ${^{16}}$O and ${^{4}}$He, which has been a
persistent problem~\cite{Pie92}. We believe the Argonne $v_{18}$ potential
has an promising future for use in microscopic nuclear many-body theory.

\acknowledgements
We wish to thank J.J.\ de Swart, J.L.\ Friar, T.-S.H.\ Lee,
V.R.\ Pandharipande, B.\ Pudliner, and S.C.\ Pieper for many useful comments
and stimulating discussions.
During the early part of this project, RBW enjoyed the hospitality and
support of S.E.\ Koonin and the Kellogg Radiation Lab of the California
Institute of Technology.
The work of RBW is supported by the U.S.\ Department of Energy, Nuclear
Physics Division, under contract No.\ W-31-109-ENG-38. The work of VGJS
was supported by the Australian Research Council, while the work of
RS was supported by the U.S.\ Department of Energy.



\narrowtext
\begin{table}
\caption{Values of fundamental constants adopted in this work.}
\begin{tabular}{cdl}
  $\hbar c$        &  197.32705   & MeV\,fm          \\
  $m_{\pi^{0}}$    &  134.9739    & MeV/c$^{2}$      \\
  $m_{\pi^{\pm}}$  &  139.5675    & MeV/c$^{2}$      \\
  $M_{p}$          &  938.27231   & MeV/c$^{2}$      \\
  $M_{n}$          &  939.56563   & MeV/c$^{2}$      \\
  $\alpha^{-1}$    &  137.03599   &                  \\
  $\mu_{p}$        &    2.79285   & $\mu_{0}$        \\
  $\mu_{n}$        &  --1.91304   & $\mu_{0}$
\end{tabular}
\label{constants}
\end{table}

\mediumtext
\begin{table}
\caption{Short-range potential parameters in MeV. The asterisk denotes that
the value was computed by Eq.~(\protect\ref{constraint}) and not fit. The
three shape parameters are: $c=2.1$ fm$^{-2}$, $r_{0}=0.5$ fm, and $a=0.2$ fm.}
\begin{tabular}{ccdddd}
   Channel       & Type  &      $I$  &      $P$  &     $Q$   &  $R$ \\
\tableline
 $S=0, T=1 (pp)$ & $c$   &--11.27028 & 3346.6874 &  1859.5627$\ast$ &    0  \\
 $S=0, T=1 (np)$ & $c$   &--10.66788 & 3126.5542 &  1746.4298$\ast$ &    0  \\
 $S=0, T=1 (nn)$ & $c$   &--11.27028 & 3342.7664 &  1857.4367$\ast$ &    0  \\
 $S=0, T=1$      & $l2$  &   0.12472 &   16.7780 &     9.0972$\ast$ &    0  \\
                 &       &           &           &                  &       \\
 $S=0, T=0$      & $c$   & --2.09971 & 1204.4301 &   511.9380$\ast$ &    0  \\
                 & $l2$  & --0.31452 &  217.4559 &   117.9063$\ast$ &    0  \\
                 &       &           &           &                  &       \\
 $S=1, T=1 (pp)$ & $c$   & --7.62701 & 1815.4920 &   969.3863$\ast$ & 1847.8059
\\
 $S=1, T=1 (np)$ & $c$   & --7.62701 & 1813.5315 &   966.2483$\ast$ & 1847.8059
\\
 $S=1, T=1 (nn)$ & $c$   & --7.62701 & 1811.5710 &   967.2603$\ast$ & 1847.8059
\\
 $S=1, T=1$      & $l2$  &   0.06709 &  342.0669 &   185.4713$\ast$ &--615.2339
\\
                 & $t$   &   1.07985 &    0      & --190.0949       &--811.2040
\\
                 & $ls$  & --0.62697 &--570.5571 & --309.3605$\ast$ &  819.1222
\\
                 & $ls2$ &   0.74129 &    9.3418 &     5.0652$\ast$ &--376.4384
\\
                 &       &           &           &                  &      \\
 $S=1, T=0$      & $c$   & --8.62770 & 2605.2682 &  1459.6345$\ast$ &  441.9733
\\
                 & $l2$  & --0.13201 &  253.4350 &   137.4144$\ast$ &  --1.0076
\\
                 & $t$   &   1.485601&    0      &--1126.8359       &  370.1324
\\
                 & $ls$  &   0.10180 &   86.0658 &    46.6655$\ast$ &--356.5175
\\
                 & $ls2$ &   0.07357 &--217.5791 & --117.9731$\ast$ &   18.3935
\end{tabular}
\label{parameters}
\end{table}

\mediumtext
\begin{table}
\caption{Distribution of $\chi^{2}$ by laboratory kinetic energy of
         the Nijmegen combined partial-wave analysis~\protect\cite{St93a}
         (PWA93) and the new Argonne $v_{18}$ potential.
         $N_{pp}$ ($N_{np}$) denotes the number of $pp$ ($np$) data
         in each energy bin.}
\begin{tabular}{r@{--}lrrrrrr}
  \multicolumn{2}{c}{} &  & \multicolumn{2}{c}{$\chi^{2}(pp)$}
                       &  & \multicolumn{2}{c}{$\chi^{2}(np)$}  \\
  \multicolumn{2}{c}{Bin (MeV)} & $N_{pp}$ & PWA93 & $v_{18}$
                                & $N_{np}$ & PWA93 & $v_{18}$   \\
\tableline
   0.0&0.5  &  134 &  134.5 &  136.3 &    10 &    9.7 &   11.8 \\
   0.5&2    &   63 &   39.7 &   41.1 &     5 &    3.8 &    7.4 \\
     2&8    &   48 &   45.0 &   36.0 &    55 &   52.4 &   51.0 \\
     8&17   &  108 &  103.0 &  111.6 &   182 &  168.3 &  164.8 \\
    17&35   &   59 &   63.1 &   72.2 &   293 &  226.6 &  234.9 \\
    35&75   &  243 &  213.4 &  251.5 &   328 &  335.2 &  339.3 \\
    75&125  &  167 &  169.5 &  171.5 &   232 &  237.1 &  231.3 \\
   125&183  &  343 &  379.7 &  415.7 &   333 &  336.8 &  363.5 \\
   183&290  &  239 &  285.9 &  304.8 &   517 &  494.6 &  574.0 \\
   290&350  &  383 &  360.7 &  421.3 &   571 &  599.0 &  708.0 \\
\tableline
     0&350  & 1787 & 1794.5 & 1962.0  & 2526 & 2463.5 & 2685.8
\end{tabular}
\label{chi2}
\end{table}

\mediumtext
\begin{table}
\caption{$pp$ phase shifts in degrees. Energies are in MeV.
         The $^{1}S_{0}$ includes the full electromagnetic interaction
         ($v^{EM}(pp)$) and is with respect to electromagnetic wave functions.
         The non-$S$ waves are nuclear phase shifts of the Coulomb
         interaction including the form factor ($V_{C1}$) with respect
         to Coulomb wave functions.}
\begin{tabular}{rdddddddd}
 $T_{\rm lab}$ & $^{1}S_{0}$ & $^{1}D_{2}$ & $^{3}P_{0}$ & $^{3}P_{1}$
               & $^{3}P_{2}$ & $\varepsilon_{2}$ & $^{3}F_{2}$ & $^{3}F_{3}$ \\
   \tableline
   1 &  32.68 &   0.00 &   0.14 & --0.08 &   0.01 & --0.00 &   0.00 &--0.00 \\
   5 &  54.74 &   0.04 &   1.61 & --0.90 &   0.22 & --0.05 &   0.00 &--0.01 \\
  10 &  55.09 &   0.17 &   3.80 & --2.05 &   0.66 & --0.20 &   0.01 &--0.03 \\
  25 &  48.51 &   0.71 &   8.78 & --4.89 &   2.49 & --0.83 &   0.10 &--0.23 \\
  50 &  38.78 &   1.73 &  11.75 & --8.23 &   5.79 & --1.77 &   0.32 &--0.69 \\
 100 &  25.01 &   3.84 &   9.61 &--13.11 &  10.98 & --2.78 &   0.73 &--1.47 \\
 150 &  15.00 &   5.77 &   4.72 &--17.27 &  14.14 & --3.02 &   1.06 &--1.96 \\
 200 &   6.99 &   7.37 & --0.50 &--21.16 &  15.91 & --2.88 &   1.24 &--2.25 \\
 250 &   0.23 &   8.61 & --5.50 &--24.86 &  16.77 & --2.58 &   1.21 &--2.45 \\
 300 & --5.64 &   9.52 &--10.17 &--28.37 &  17.01 & --2.23 &   0.90 &--2.66 \\
 350 &--10.86 &  10.14 &--14.49 &--31.70 &  16.81 & --1.88 &   0.29 &--2.95
\end{tabular}
\label{phspp}
\end{table}

\mediumtext
\begin{table}
\caption{$nn$ phase shifts in degrees. Energies are in MeV.
         The $^{1}S_{0}$ includes the full electromagnetic interaction
         ($v^{EM}(nn)$). The non-$S$ waves only include the nuclear
         interaction.
         All phase shifts are with respect to Riccati-Bessel functions.}
\begin{tabular}{rdddddddd}
 $T_{\rm lab}$ & $^{1}S_{0}$ & $^{1}D_{2}$ & $^{3}P_{0}$ & $^{3}P_{1}$
               & $^{3}P_{2}$ & $\varepsilon_{2}$ & $^{3}F_{2}$ & $^{3}F_{3}$ \\
   \tableline
   1 &  57.07 &   0.00 &   0.21 & --0.12 &   0.02 & --0.00 &   0.00 &--0.00 \\
   5 &  60.64 &   0.05 &   1.88 & --1.04 &   0.27 & --0.06 &   0.00 &--0.01 \\
  10 &  57.48 &   0.18 &   4.17 & --2.24 &   0.76 & --0.22 &   0.01 &--0.04 \\
  25 &  48.80 &   0.74 &   9.13 & --5.12 &   2.69 & --0.86 &   0.11 &--0.24 \\
  50 &  38.47 &   1.79 &  11.89 & --8.48 &   6.08 & --1.80 &   0.32 &--0.70 \\
 100 &  24.45 &   3.92 &   9.48 &--13.38 &  11.31 & --2.79 &   0.74 &--1.49 \\
 150 &  14.38 &   5.87 &   4.46 &--17.58 &  14.45 & --3.00 &   1.07 &--1.98 \\
 200 &   6.34 &   7.48 & --0.81 &--21.49 &  16.19 & --2.84 &   1.25 &--2.26 \\
 250 & --0.42 &   8.72 & --5.85 &--25.21 &  17.00 & --2.53 &   1.21 &--2.46 \\
 300 & --6.31 &   9.62 &--10.54 &--28.73 &  17.20 & --2.18 &   0.88 &--2.68 \\
 350 &--11.53 &  10.24 &--14.87 &--32.08 &  16.96 & --1.83 &   0.25 &--2.97
\end{tabular}
\label{phsnn}
\end{table}

\mediumtext
\begin{table}
\caption{$np$ $T=1$ phase shifts in degrees. Energies are in MeV.
         The $^{1}S_{0}$ includes the full electromagnetic interaction
         ($v^{EM}(np)$). The non-$S$ waves only include the nuclear
         interaction.
         All phase shifts are with respect to Riccati-Bessel functions.}
\begin{tabular}{rdddddddd}
 $T_{\rm lab}$ & $^{1}S_{0}$ & $^{1}D_{2}$ & $^{3}P_{0}$ & $^{3}P_{1}$
               & $^{3}P_{2}$ & $\varepsilon_{2}$ & $^{3}F_{2}$ & $^{3}F_{3}$ \\
\tableline
   1 &  62.02 &   0.00 &   0.18 & --0.11 &   0.02 & --0.00 &   0.00 &--0.00 \\
   5 &  63.50 &   0.04 &   1.64 & --0.93 &   0.26 & --0.05 &   0.00 &--0.00 \\
  10 &  59.78 &   0.16 &   3.71 & --2.04 &   0.72 & --0.19 &   0.01 &--0.03 \\
  25 &  50.61 &   0.68 &   8.32 & --4.82 &   2.57 & --0.77 &   0.08 &--0.20 \\
  50 &  40.09 &   1.70 &  10.99 & --8.15 &   5.86 & --1.68 &   0.28 &--0.61 \\
 100 &  26.02 &   3.81 &   8.69 &--13.07 &  11.00 & --2.69 &   0.67 &--1.35 \\
 150 &  15.98 &   5.72 &   3.78 &--17.28 &  14.12 & --2.95 &   0.98 &--1.82 \\
 200 &   8.00 &   7.30 & --1.43 &--21.22 &  15.86 & --2.82 &   1.15 &--2.10 \\
 250 &   1.28 &   8.52 & --6.41 &--24.95 &  16.70 & --2.54 &   1.10 &--2.30 \\
 300 & --4.54 &   9.43 &--11.06 &--28.49 &  16.91 & --2.21 &   0.77 &--2.51 \\
 350 & --9.71 &  10.06 &--15.36 &--31.85 &  16.69 & --1.88 &   0.14 &--2.81
\end{tabular}
\label{phsnp1}
\end{table}

\mediumtext
\begin{table}
\caption{$np$ $T=0$ phase shifts in degrees. Energies are in MeV.
         The coupled $^{3}S_{1}$-$^{3}D_{1}$ channel includes the full
         electromagnetic interaction ($v^{EM}(np)$).
         The non-$S$ waves only include the nuclear interaction.
         All phase shifts are with respect to Riccati-Bessel functions.}
\begin{tabular}{rddddddddd}
 $T_{\rm lab}$ & $^{1}P_{1}$ & $^{1}F_{3}$ & $^{3}S_{1}$
               & $\varepsilon_{1}$ & $^{3}D_{1}$ & $^{3}D_{2}$
               & $^{3}D_{3}$ & $\varepsilon_{3}$ & $^{3}G_{3}$ \\
\tableline
   1 & --0.19 &--0.00 & 147.75 & 0.11 & --0.00 &  0.01 & 0.00 & 0.00 &--0.00 \\
   5 & --1.51 &--0.01 & 118.18 & 0.66 & --0.17 &  0.22 & 0.00 & 0.01 &--0.00 \\
  10 & --3.11 &--0.07 & 102.62 & 1.14 & --0.65 &  0.85 & 0.01 & 0.08 &--0.00 \\
  25 & --6.48 &--0.42 &  80.68 & 1.77 & --2.72 &  3.71 & 0.08 & 0.55 &--0.05 \\
  50 & --9.85 &--1.13 &  62.89 & 2.11 & --6.28 &  8.94 & 0.40 & 1.61 &--0.26 \\
 100 &--14.20 &--2.22 &  43.51 & 2.52 &--12.04 & 17.10 & 1.61 & 3.50 &--0.93 \\
 150 &--17.68 &--2.98 &  31.19 & 2.96 &--16.39 & 21.85 & 2.92 & 4.88 &--1.74 \\
 200 &--20.79 &--3.61 &  21.94 & 3.43 &--19.82 & 24.20 & 4.00 & 5.88 &--2.58 \\
 250 &--23.65 &--4.22 &  14.45 & 3.92 &--22.59 & 25.06 & 4.76 & 6.61 &--3.41 \\
 300 &--26.28 &--4.87 &   8.13 & 4.43 &--24.83 & 25.01 & 5.21 & 7.16 &--4.20 \\
 350 &--28.71 &--5.59 &   2.65 & 4.95 &--26.65 & 24.41 & 5.39 & 7.59 &--4.96
\end{tabular}
\label{phsnp0}
\end{table}


\narrowtext
\begin{table}
\caption{Scattering lengths and effective ranges in fm.}
\begin{tabular}{cd@{$\pm$}rdd}
  & \multicolumn{2}{c}{Experiment} & Argonne $v_{18}$ & w/o $v^{EM}$  \\
\tableline
  $^{1}a_{pp}$ &  --7.8063 & 0.0026$^{\rm a}$ &  --7.8064 & --17.164  \\
  $^{1}r_{pp}$ &    2.794  & 0.014$^{\rm a}$  &    2.788  &    2.865  \\
  $^{1}a_{nn}$ & --18.5    & 0.5$^{\rm b}$    & --18.487  & --18.818  \\
  $^{1}r_{nn}$ &    2.8    & 0.1$^{\rm b}$    &    2.840  &    2.834  \\
  $^{1}a_{np}$ & --23.749  & 0.008$^{\rm c}$  & --23.732  & --23.084  \\
  $^{1}r_{np}$ &    2.81   & 0.05$^{\rm c}$   &    2.697  &    2.703  \\
  $^{3}a_{np}$ &    5.424  & 0.003$^{\rm c}$  &    5.419  &    5.402  \\
  $^{3}r_{np}$ &    1.760  & 0.005$^{\rm c}$  &    1.753  &    1.752
\end{tabular}
\tablenotetext[1]{Ref.~\protect\cite{Ber88}}
\tablenotetext[2]{Ref.~\protect\cite{Gab79}}
\tablenotetext[3]{Ref.~\protect\cite{Koe75}}
\label{lowpar}
\end{table}

\narrowtext
\begin{table}
\caption{Evolution of $^{1}S_{0}$ $pp$ phase shifts from the
         charge-independent potential to the full interaction,
         as described in the text. Energies are in MeV.}
\begin{tabular}{rddddd}
 $T_{\rm lab}$ & CI & + $m_{p}$ & + CD $v^{\pi}$ & + CD $v^{R}$ & + $v^{EM}$ \\
\tableline
   1 &  58.40 &  57.20 &  57.42 &  55.50 &  32.68 \\
   5 &  61.44 &  61.34 &  60.88 &  59.78 &  54.74 \\
  10 &  58.14 &  58.07 &  57.71 &  56.84 &  55.09 \\
  25 &  49.33 &  49.29 &  49.05 &  48.36 &  48.51 \\
  50 &  38.95 &  38.93 &  38.76 &  38.13 &  38.78 \\
 100 &  24.93 &  24.92 &  24.80 &  24.19 &  25.01 \\
 150 &  14.87 &  14.87 &  14.77 &  14.16 &  15.00 \\
 200 &   6.86 &   6.86 &   6.77 &   6.15 &   6.99 \\
 250 &   0.11 &   0.12 &   0.04 & --0.60 &   0.23 \\
 300 & --5.75 & --5.74 & --5.82 & --6.47 & --5.64 \\
 350 &--10.96 &--10.95 &--11.01 &--11.69 &--10.86
\end{tabular}
\label{phsevol}
\end{table}

\narrowtext
\begin{table}
\caption{Static deuteron properties.}
\begin{tabular}{cdddl}
               &  Experiment  &  Argonne $v_{18}$  & + R + MEC  &  Units   \\
\tableline
   $E_{d}$     & 2.224575(9)$^{\rm a}$ &    2.224575 &       &  MeV        \\
   $<T>$       &                       &   19.814    &       &  MeV        \\
   $<v^{EM}>$  &                       &    0.018    &       &  MeV        \\
   $<v^{\pi}>$ &                       & --21.286    &       &  MeV        \\
   $<v^{R}>$   &                       &  --0.770    &       &  MeV        \\
   $A_{S}$     & 0.8846(8)$^{\rm b}$   &    0.8850   &       &  fm$^{1/2}$ \\
   $\eta$      & 0.0256(4)$^{\rm c}$   &    0.0250   &       &             \\
   $r_{d}$     & 1.9660(68)$^{\rm d}$  &    1.967    &       &  fm         \\
   $\mu_{d}$   & 0.857406(1)$^{\rm e}$ &    0.847    & 0.871 & $\mu_{0}$   \\
   $Q_{d}$     & 0.2859(3)$^{\rm f}$   &    0.270    & 0.275 &  fm$^{2}$   \\
   $P_{d}$     &                       &    5.76     &       &  \%
\end{tabular}
\tablenotetext[1]{Ref.~\protect\cite{Leu82}}
\tablenotetext[2]{Ref.~\protect\cite{Eri83}}
\tablenotetext[3]{Ref.~\protect\cite{Rod90}}
\tablenotetext[4]{Ref.~\protect\cite{Sim81}}
\tablenotetext[5]{Ref.~\protect\cite{Lin65}}
\tablenotetext[6]{Ref.~\protect\cite{Bis79}}
\label{deuteron}
\end{table}


\begin{figure}
\caption{Form factors in the electromagnetic interaction.}
\label{figFFs}
\end{figure}

\begin{figure}
\caption{Phase shifts in the $^{1}S_{0}$ channel for $np$, $nn$, and
$pp$ scattering, compared to various partial-wave phase-shift analyses.}
\label{fig1S0}
\end{figure}

\begin{figure}
\caption{Phase shifts in the $^{3}P_{0}$ channel for $np$, $nn$, and
$pp$ scattering, compared to various partial-wave phase-shift analyses.}
\label{fig3P0}
\end{figure}

\begin{figure}
\caption{The $\epsilon_{1}$ mixing parameter compared to various
partial-wave phase-shift analyses.}
\label{figeps1}
\end{figure}

\begin{figure}
\caption{Phase shifts in the $^{1}P_{1}$ channel, compared to various
partial-wave phase-shift analyses.}
\label{fig1P1}
\end{figure}

\begin{figure}
\caption{Central, isospin, spin, and spin-isospin components of the
potential.  The central potential has a peak value of 2031 MeV at $r=0$.}
\label{figvc}
\end{figure}

\begin{figure}
\caption{Tensor and tensor-isospin parts of the potential.  Also shown
are the OPE contribution to the tensor-isospin potential, and for comparison,
an OPE potential with a monopole form factor containing a 900 MeV cutoff mass.}
\label{figvt}
\end{figure}

\begin{figure}
\caption{Spin-orbit and quadratic spin-orbit components of the potential.}
\label{figvls}
\end{figure}

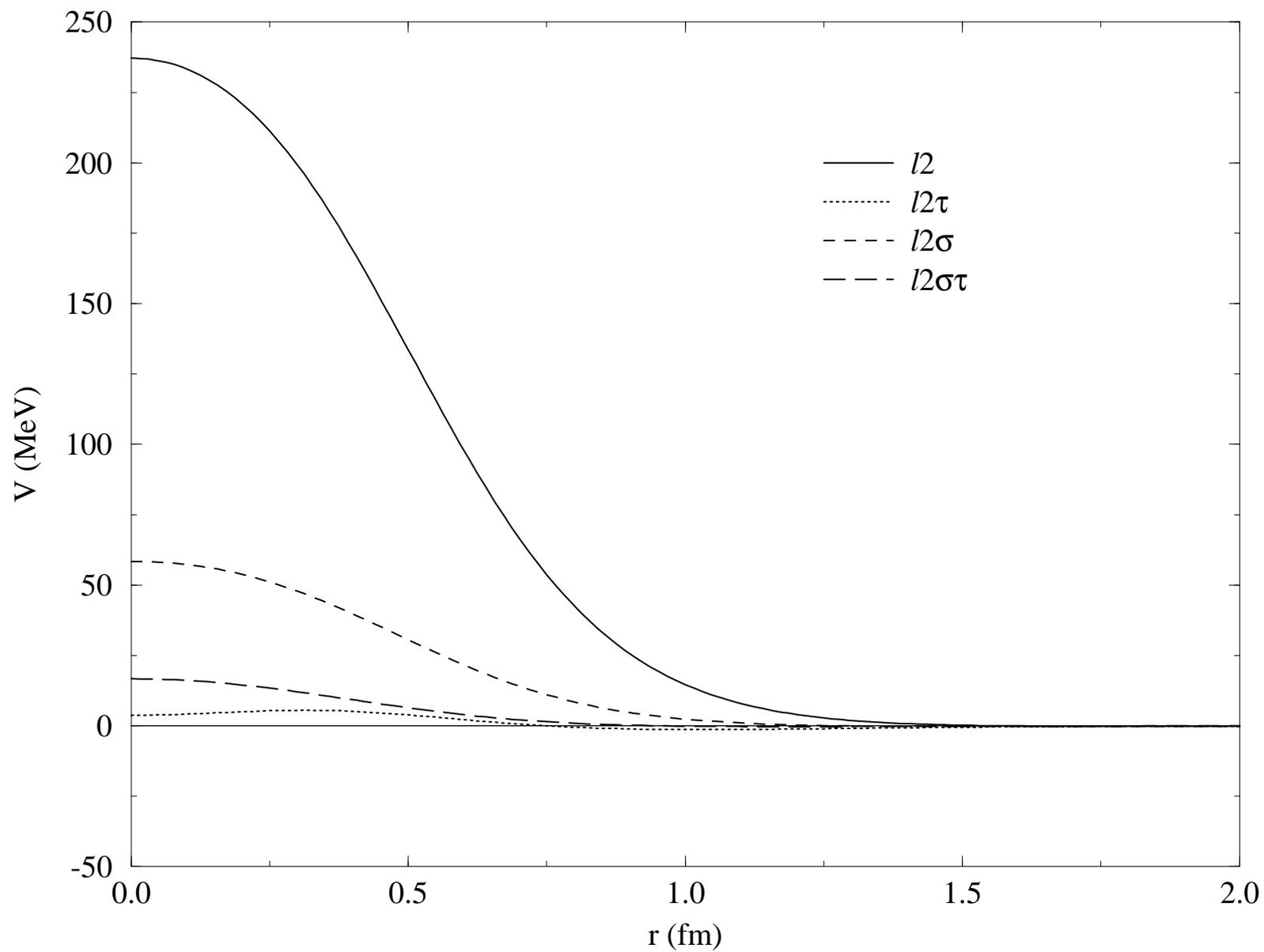
\begin{figure}
\caption{$L^{2}$ components of the potential.}
\label{figvl2}
\end{figure}

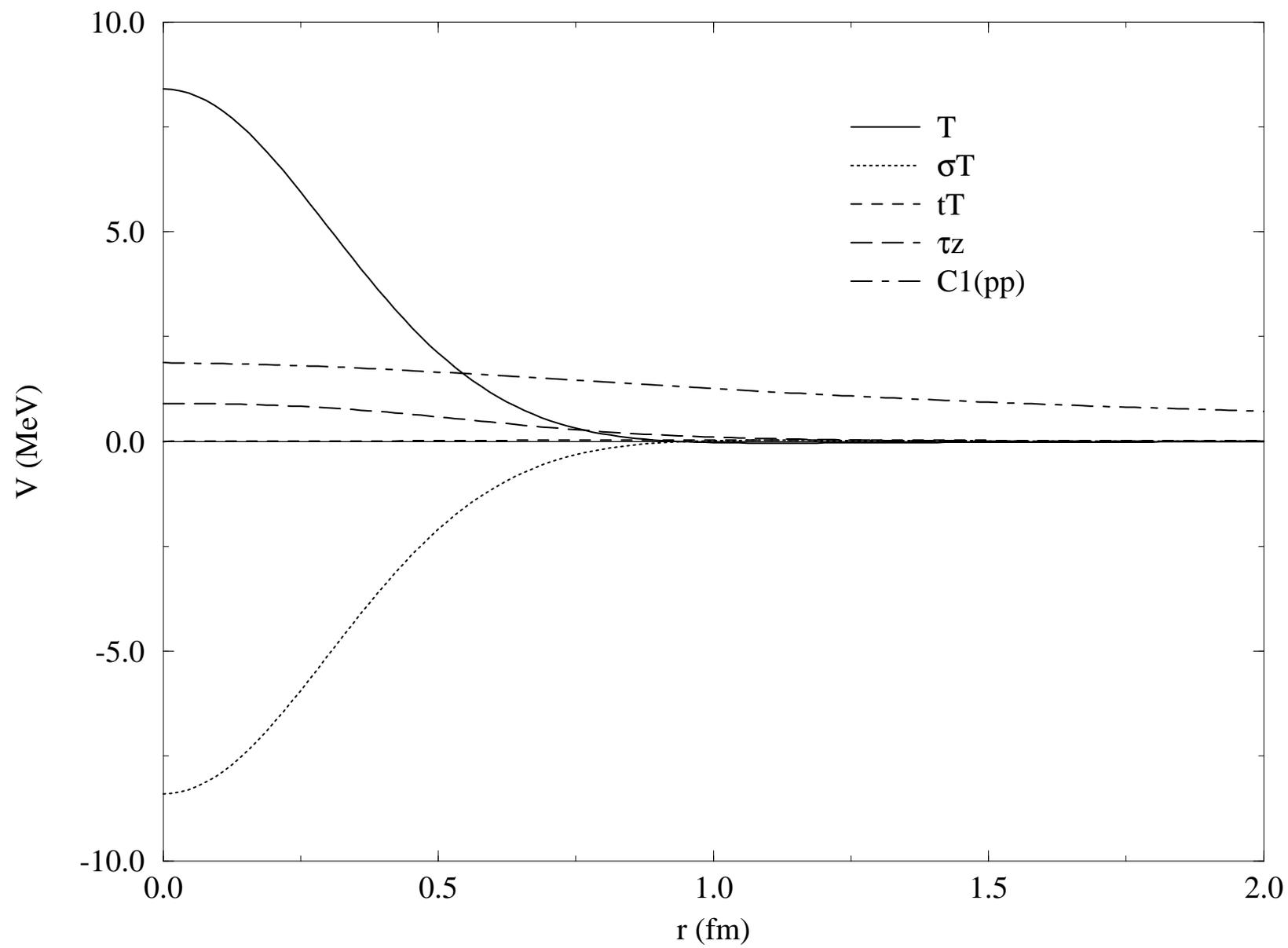
\begin{figure}
\caption{Charge-dependent and charge-asymmetric components of the potential.
Also shown for comparison is the Coulomb potential, $V_{C1}(pp)$.}
\label{figvcd}
\end{figure}

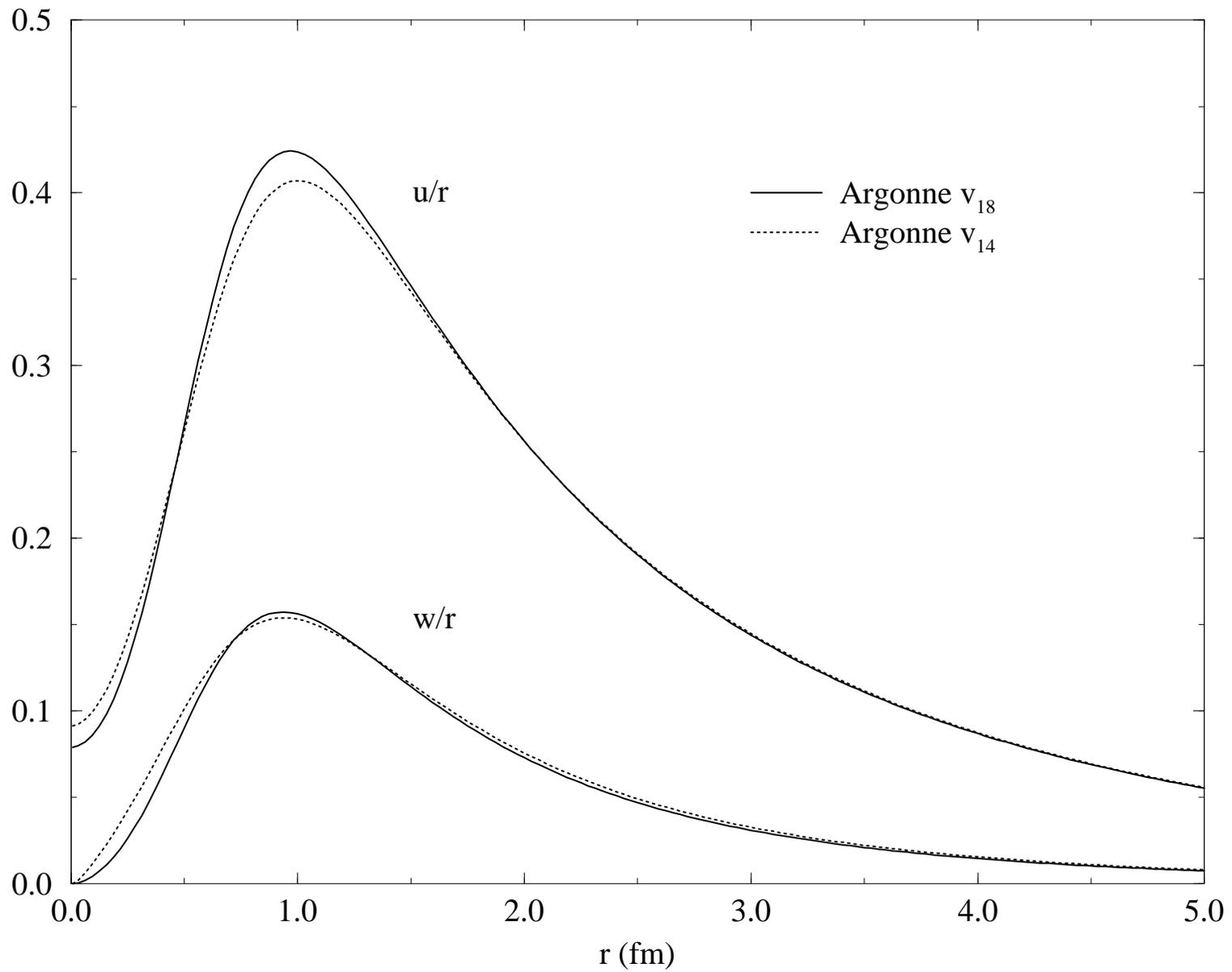
\begin{figure}
\caption{The deuteron $S$- and $D$-wave function components divided by $r$.}
\label{figuw}
\end{figure}

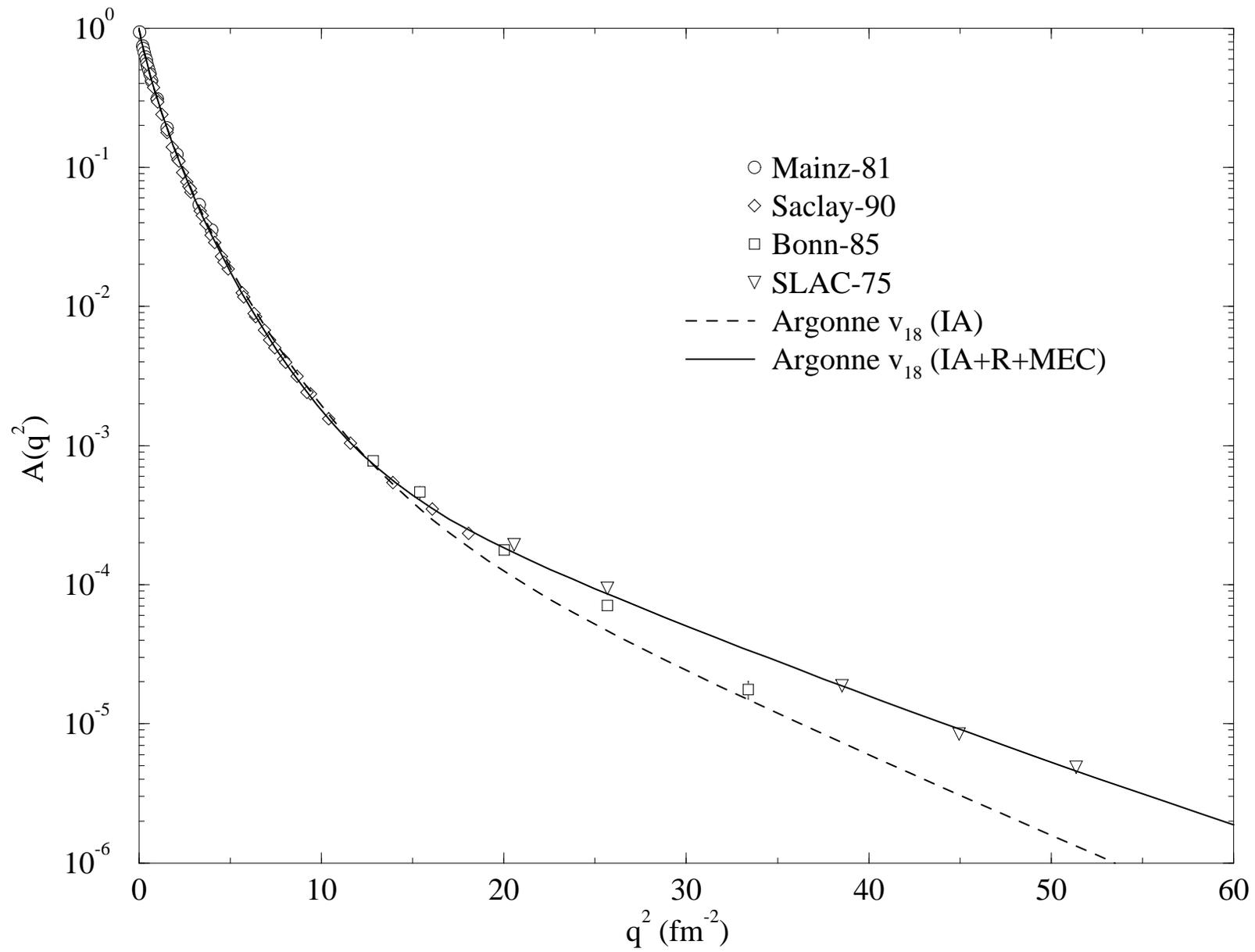
\begin{figure}
\caption{The deuteron electromagnetic structure function $A(q^{2})$ in impulse
approximation (dashed line) and with relativistic and exchange-current
corrections (solid line).
Data are from Bonn~\protect\cite{Cra85}, Mainz~\protect\cite{Sim81},
Saclay~\protect\cite{Pla90}, and SLAC~\protect\cite{Arn75}.}
\label{figA}
\end{figure}

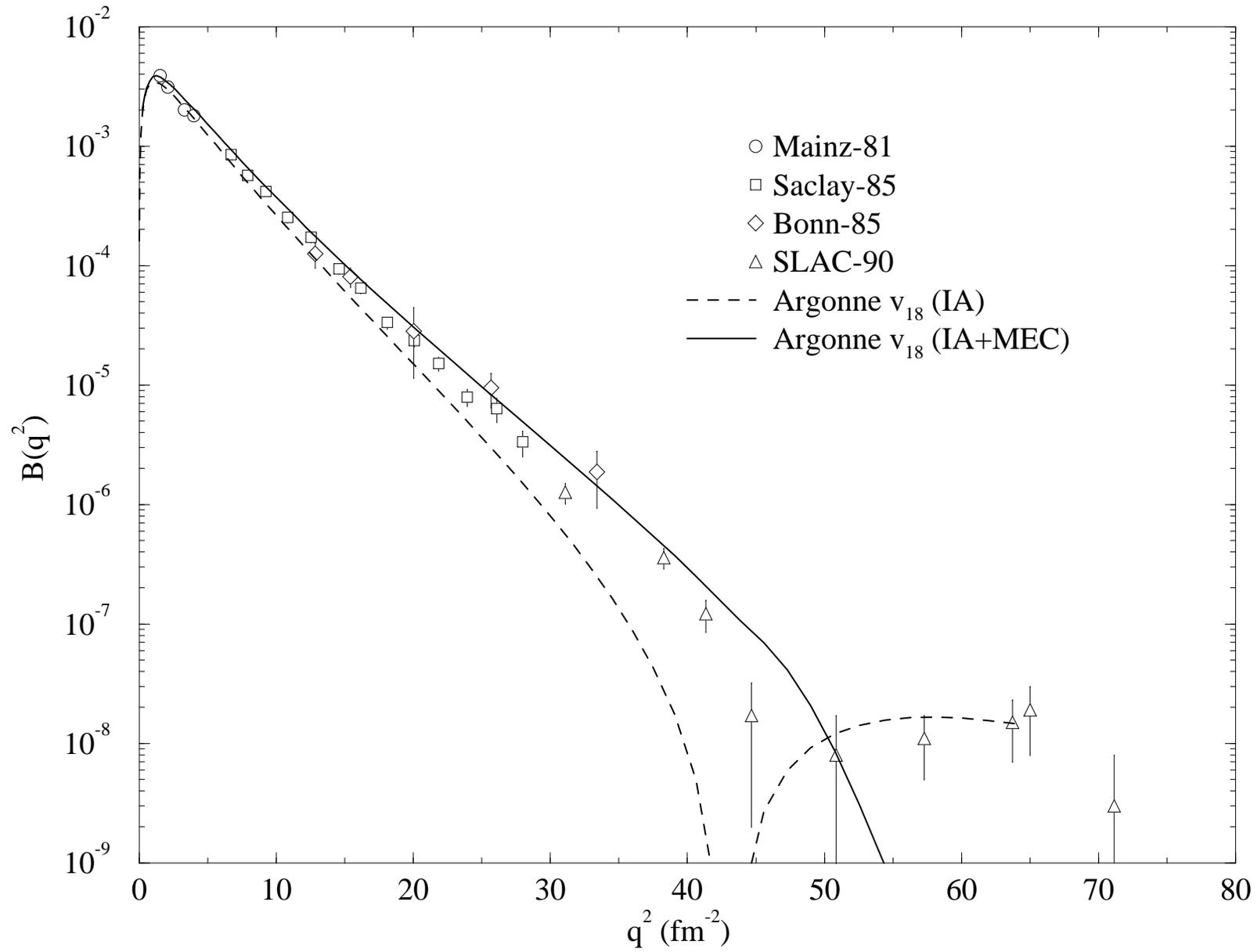
\begin{figure}
\caption{The deuteron electromagnetic structure function $B(q^{2})$ in impulse
approximation (dashed line) and with exchange-current corrections (solid line).
Data are from Bonn~\protect\cite{Cra85}, Mainz~\protect\cite{Sim81},
Saclay~\protect\cite{Auf85}, and SLAC~\protect\cite{Arn87}.}
\label{figB}
\end{figure}

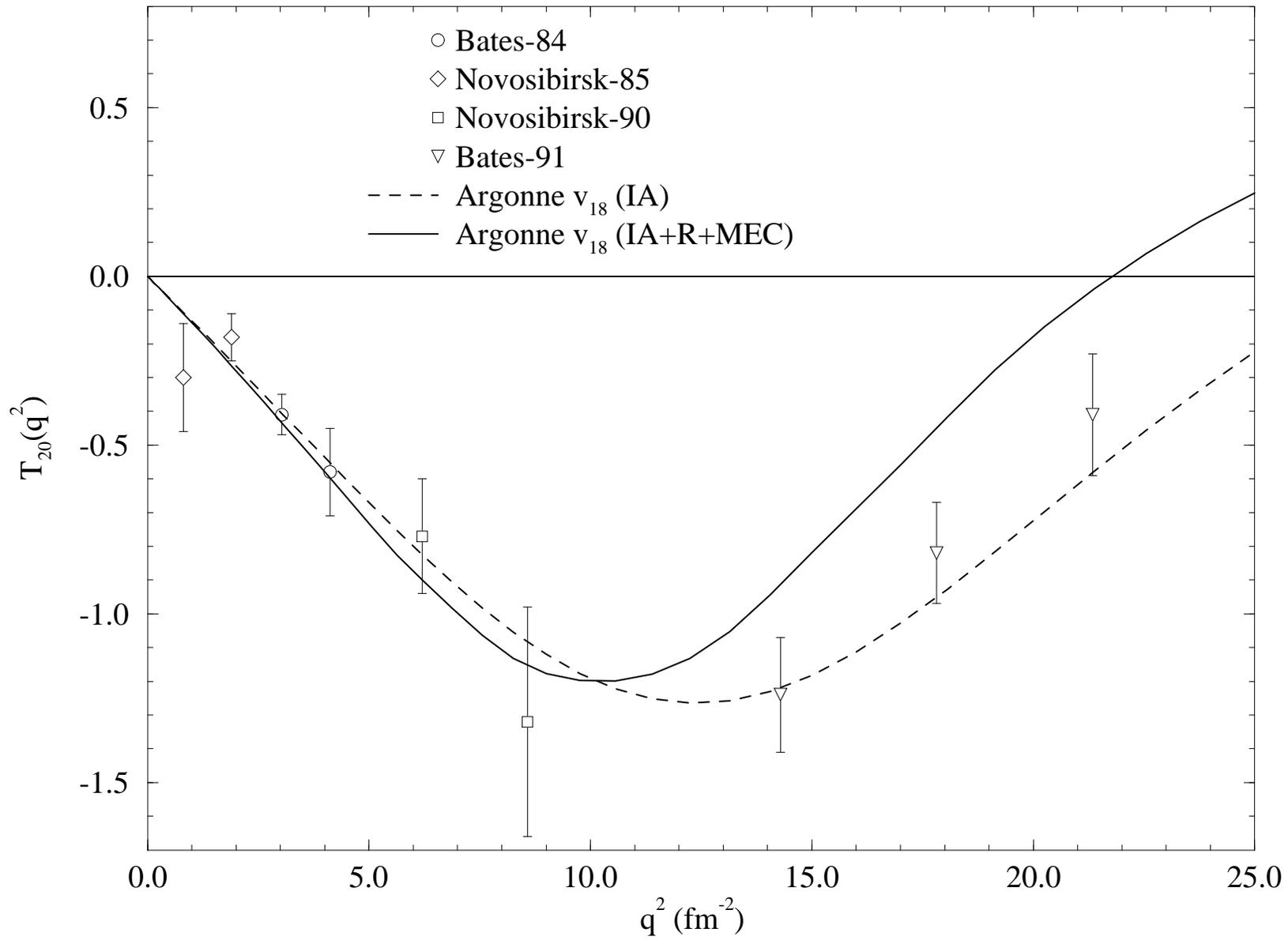
\begin{figure}
\caption{The deuteron tensor polarization $T_{20}$ in impulse approximation
(dashed line) and with relativistic and exchange-current corrections
(solid line).  Data are from Bates~\protect\cite{Sch84,The91} and
Novosibirsk~\protect\cite{Dmi85,Gil90}.}
\label{figT20}
\end{figure}

\end{document}